\documentclass[a4paper]{article}

\usepackage{INTERSPEECH2020}
\usepackage{cite}
\usepackage{graphicx}
\usepackage{float}

\title{The ASRU 2019 Mandarin-English Code-Switching Speech Recognition Challenge: Open Datasets, Tracks, Methods and Results}
\name{Xian Shi$^1$, Qiangze Feng$^2$, Lei Xie$^1$}
\address{
  $^1$Audio, Speech and Language Processing Group (ASLP@NPU), School of Computer Science, Northwestern Polytechnical University, Xi'an, China\\
  $^2$Datatang (Beijing) Technology Co., LTD, Beijing, China}
\email{xshi@npu-aslp.org, xiaoqiang@datatang.com, lxie@nwpu.edu.cn}

\begin{document}
	\maketitle
	\begin{abstract}
		Code-switching (CS) is a common phenomenon and recognizing CS speech is challenging. But CS speech data is scarce and there's no common testbed in relevant research. This paper describes the design and main outcomes of the ASRU 2019 Mandarin-English code-switching speech recognition challenge, which aims to improve the ASR performance in Mandarin-English code-switching situation. 500 hours Mandarin speech data and 240 hours Mandarin-English intra-sentencial CS data are released to the participants. Three tracks were set for advancing the AM and LM part in traditional DNN-HMM ASR system, as well as exploring the E2E models' performance.
		The paper then presents an overview of the results  and system performance in the three tracks. It turns out that traditional ASR system benefits from pronunciation lexicon, CS text generating and data augmentation. In E2E track, however, the results highlight the importance of using language identification, building-up a rational set of modeling units and spec-augment. The other details in model training and method comparsion are discussed.
	\end{abstract}
	
	\section{Introduction}
	
	Code-Switching (CS), the alternating use of more than one languages inside a single utterance\cite{cs}, is a special and complicated language phenomenon, which has become an important field of both linguistics and ASR research. For instance, the Interspeech 2020 workshop on speech technologies for code switching is a recent platform particularly focusing on CS related research.\footnote{\emph{www.microsoft.com/en-us/research/event/workshop-on-speech-technologies-for-code-switching-2020}}
	Code-switching has many varieties, and a classification method based on mixed position is often used, which classifies CS into two primary categories: \emph{inter-sentencial} (switch happens at the sentence boundaries) and \emph{intra-sentencial} (switch happens in the middle of a sentence).
	%In recent years, CS appears more frequently in world wide conversations such as Mandarin-English, Hindi-English and Spanish-English.
	
	An ASR system usually contains the ability of modeling linguistic information and acoustic information at the same time. In CS situation however, language switching happens at unpredictable positions makes it difficult to train a multilingual language model (LM) while the varied accent of non-native speakers and mixing of phonemes from different language bring difficulty to acoustic model (AM) training. 
	
	Previous work has made continous progress in Code-switching area, and a variety of modeling methods are proposed, which can be roughly divided into three categories.
	The first kind of optimization aims at modeling units, which turns out that both phone merging methods and new modeling units building methods \cite{baayn, lin2009study} are helpful to code-switching ASR. The second kind of methods focus on the nerual network structure, making deep neural networks-hidden Markov model (DNN-HMM) based ASR system more competent for CS tasks by optimizing the neural network and training strategy\cite{tong2017investigation, ghoshal2013multilingual}. The third kind of efforts is explored on End-to-End (E2E) speech recognition\cite{vaswani2017attention, chan2016listen, hori2017advances, watanabe2017hybrid}. E2E ASR framework enables lexicon-free recognition, which is an important advantage over traditional hybrid system, especially for CS tasks. 
	An encoder-decoder based CS ASR system was built by Hiroshi et al.\cite{seki2018end}.
	Zhang et al.\cite{zhang2019towards} built a bilingual Mandarin-English acoustic model by putting two separately pre-trained DFSMN-CTC-sMBR together.
	Li et al.\cite{li2019towards} added a frame-level language identification (LID) loss to bilingual CTC model, assisting CTC to distinguish the language ID of frames.
	Although those efforts have improved the performance of CS ASR, robust ASR system that supports arbitrary switching of languages still remains a challenging goal.
	
	This paper describes the design and outcomes of the ASRU 2019 Mandarin-English Code-Switching Speech Recognition Challenge, a special event of IEEE Automatic Speech Recognition and Understanding Workshop (ASRU 2019).Actually, the difficulties discussed above also reveal one of the bottelnecks in CS ASR research: \emph{data insufficiency}. Code-switching speech data is always scarce, only SEAME\cite{SEAME}, a small set of 30 hours Mandarin-English speech data collected in Singapore and Malaysia is released to the public. Besides, in Mandarin-English CS ASR, there is no common testbed and open datasets for method validation and model comparison, especially in the area of fast development of data-hungry deep learning approaches. This challenge is especially designed for these reasons. Three speech datasets are released to participants, 740 hours in total, and 240 hours of them are Mandarin-English CS data\footnote{Exploring \emph{www.datatang.com/competition} for more detail about datasets and the challenge.}. A well-trained 3-gram CS language model in ARPA format is also provided. The participants are supposed to use the permitted data only to build CS ASR systems in three tracks: i) Traditional ASR system with identical official N-gram LM; ii) Traditional ASR system without LM limitation; iii) End-to-End ASR system. Totally 72 teams participated in the challenge. Participants have around 50 days to finish their system building and submit the recognition results.
	
	The rest of this paper is organized as below:  Section 2 describes detailed information of the datasets. In Section 3, rules, data using limitation of each track and results evaluation method are explained. Section 4 describes a overview of results submitted and advancing system building methods in all the tracks. Summary of the main findings in the challenge is in Section 5.
	
	\section{Open Source Datasets}
	
	Code-switching speech data is always scarce, which hinders the research of CS ASR seriously. For this challenge, DataTang released 3 ASR datasets to participants. The basic information of the datasets is as below.
	
	\begin{table}[h]
		\centering
		\caption{Basic information of the 3 released datasets}
		\begin{tabular}{lll}
			\hline
			Dataset   & Transcripts Type                      & Dur/hours  \\ \hline
			\emph{Train$_{Man}$} & Mandarin only                        & 500 \\
			\emph{Train$_{CS}$}  & Intra-sen Mandarin-English CS & 200 \\
			\emph{Dev$_{CS}$}    & Intra-sen Mandarin-English CS & 40  \\ \hline
		\end{tabular}
		\label{data}
	\end{table}
	
	All the data are collected by smart phones in quiet rooms from various Android phones and iPhones. The speakers were from 30 provinces in China. 70\% of the speakers were under 30 years old, with no significant difference in the number of male and female.
	
	The transcripts of data cover many common fields including entertainment, travel, daily life and social interaction. In \emph{Train$_{Man}$}, each sentence has 10 Chinese characters in average. As for \emph{Train$_{CS}$} and \emph{Dev$_{CS}$}, each sentence has 8.6 Chinese characters and 1.6 English words in average. Most English words are nouns, personal names, song names and some adjectives. Besides, there are 6 kinds of symbols and tags for noise and English abbreviation in \emph{Train$_{CS}$} and \emph{Dev$_{CS}$} transcripts. Several examples from the dataset are shown in Table \label{texteg}.
	
	\begin{table}[h]
		\centering
		\footnotesize
		\caption{Examples of CS transcription from the dataset}
		\includegraphics[width=\linewidth,scale=1.00]{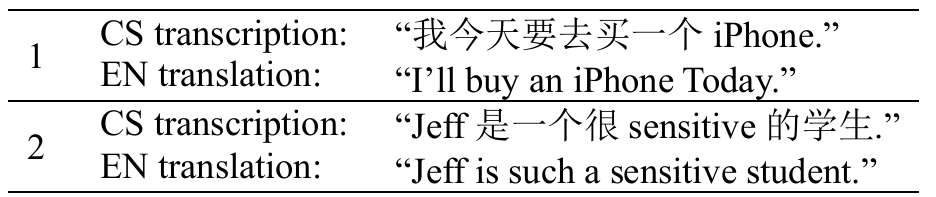} 
		\label{texteg}
		\vspace{-0.35cm}
	\end{table}
		
%	, shown in Table \ref{tag}.
%	
%	\begin{table}[h]
%		\centering
%		\caption{Symbles and tags in transcripts.}
%		\begin{tabular}{ll}
%			\hline
%			Symbol\&Tag             & Indication                           \\ \hline
%			Noise symbol 1: {[}S{]} & Non text noise from speaker          \\
%			Noise symbol 2: {[}N{]} & Inhuman sound like knocking keyboard \\
%			Noise symbol 3: {[}T{]} & Stable noise like wind               \\
%			Noise symbol 4: {[}P{]} & Non text noise from other people     \\
%			Abbreviation tag: !     & Pronounce by a word                  \\
%			Abbreviation tag: @     & Spell                                \\ \hline
%		\end{tabular}
%		\label{tag}
%	\end{table}
	
	\section{Tracks Setting and Rules}
	
	Traditional forced alignment based ASR system consists of two separately trained components: acoustic model and language model. The hybrid MMI-Chain model build by Kaldi\cite{povey2011kaldi} is regarded as one of the state-of-the-art ASR systems. The first two tracks aim at making the traditional ASR system able to recognize Mandarin-English code-switching speech. E2E ASR is drawing increasing attention in recent years, and it seems to have more potential and possibility to solve CS question. Track 3 was built for E2E systems to contest with each other.
	
	A series of instructions are designed to ensure an equitable comparison. In Section 3.4, the evaluation of code-switching recognition results is explained.
	
	\subsection{Track1: Traditional ASR with identical N-gram LM}
	
	Acoustic model in traditional ASR system is used to bind a speech frame to a certain unit through computing acoustic likelihood. This track focus on AM behavior only. 
	
	A Mandarin language model and a code-switching language model are trained separately by KenLM toolkit\cite{heafield2011kenlm} and merged with SRILM toolkit\cite{stolcke2002srilm}. The triple-gram in Mandarin training data occurs less than 10 times are pruned while no pruning was applied to uni-gram and bi-gram. The final size of the merged arpa is 2G.
	
	Rules that participants should follow in track 1 is as below:
	\begin{enumerate}
		\item The acoustic model should be a frame-level force-alignment model, and CTC model is prohibited.
		\item Data used for AM training is limited to \emph{Train$_{Man}$}, \emph{Train$_{CS}$} and Librispeech\cite{panayotov2015librispeech} 960 hours English speech data. Data augmentation methods are allowed.
%		\item No pronunciation lexicon and phone set is provided, participants should finish transcripts pre-processing and word segment by themselves.
		\item Multi-system fusion techniques including recognizer output voting error reduction (ROVER)\cite{fiscus1997post} are prohibited.
		\item Decoding graph should be complied with G.fst generated by the given arpa. Any kind of lattice rescoring is prohibited.
	\end{enumerate}
	
	\subsection{Track2: Traditional ASR without LM limitation}
	
	Language model also plays an important role in traditional ASR. It estimates the grammatical rationality of character or word sequences. In this track, any training data for LM and all kinds of techniques including but not limited to RNN LM , large scale LM rescoring are allowed, but AM still should be trained under the rule 1-3 of track 1.
	
%		\begin{table}[b]
%			\caption{Edit Distence in MER calculation}
%			\centering
%			\label{MER}
%			\begin{tabular}{llllll}
%				\hline
%				\textbf{ref} & \textit{Maybe} & \textit{他} & \textit{去} & \textit{Singapore} & \textit{-} \\
%				\textbf{dec} & \textit{-}     & \textit{我} & \textit{去} & \textit{新}         & \textit{坡} \\
%				& $\mid$              & $\mid$          & $\mid$          & $\mid$                  & $\mid$          \\
%				& en-del         & cn-sub     & cn-cor        & en-sub             & cn-ins     \\ \hline
%				\multicolumn{6}{l}{MER/CH ER/EN ER: 80\%, 100\%, 33\%}                                  \\ \hline
%			\end{tabular}
%		\end{table}
	
	\subsection{Track3: End-to-End ASR}
	
	End-to-End ASR here refers to systems without frame-level forced-alignment, always modeling acoustic information and language information jointly. It is becoming an increasingly topical field and various of E2E ASR systems are proposed. Encoder-Decoder based system LAS\cite{chan2016listen} and transformer\cite{vaswani2017attention} use global attention and multi-head self-attention to generate implicit alignment. RNN-transducer combines two RNNs into a sequence transduction system\cite{graves2012sequence, boulanger2013high}.
	
	Including the models above, any E2E ASR system is allowed in track 3, and CTC model is also regarded as an E2E model. Rule 2-3 in track 1 are effective in track 3. Besides, as for systems need to model acoustic information and language information jointly, the text training data is limited to transcripts of permitted speech data.
	
	\begin{table*}[htbp] 
		\centering
		\caption{Top 10 of 35 submitted systems in track 1. The columns in the middle summarize the key features of systems, CER(\%) for Chinese part error rate, WER(\%) for English part error rate, and MER(\%) for mixture. The right side of the table describe the error rate of Mandarin part, English part and total MER. Phone merging includes partial combining and totally binding of Chinese phones and English phones.}
		\centering 
		\includegraphics[width=\linewidth,scale=1.00]{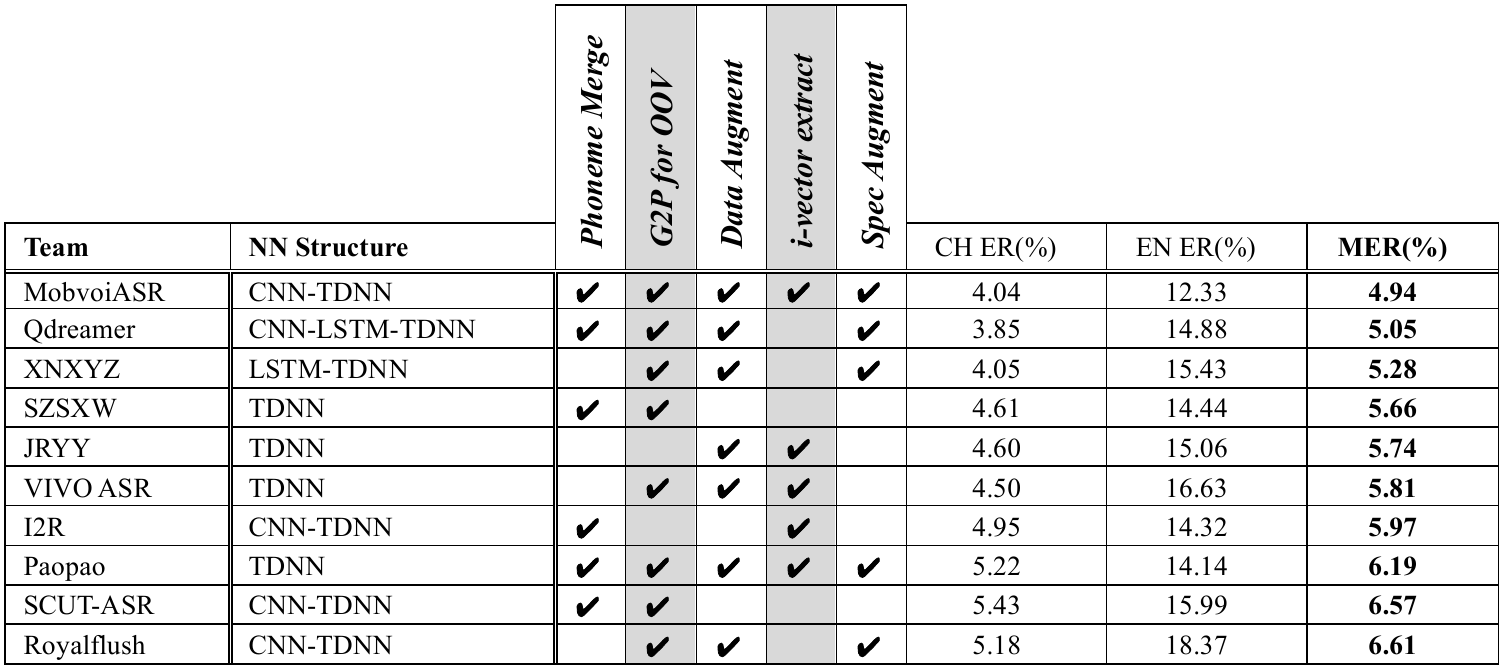} 
		\label{track1}
		\vspace{-0.5cm}
	\end{table*}

	\subsection{Results Submission and Evaluation Plan}
	
	Competitors are supposed to submit their recognition results and system descriptions of each track they participated in. Recognition accuracy is the only target considered in the evaluation. Mixture error rate (MER) considers Mandarin characters and English words as the tokens in the edit distance calculation. Errors of Chinese and English will be counted separately according to the language of the reference token. 
	%A simple example of MER calculation is in Table \ref{MER}.

	The error rate of the Chinese part and the English part in the final publicity result is only for reference, ranking is based on MER only.

	\section{Results and Discussion on Methods}
	
	\subsection{Track1}
	
	35 teams submit their results of track 1, the top 10 best systems is listed in Table \ref{track1}, along with the their key features. The following part introduces the main outcomes in three aspects.

	\subsubsection{Phone Sets}
%	\noindent\textbf{Phone Sets: }
	Building a traditional ASR system starts with building a phone set. 
%	There are four types of Mandarin-English mixture phone set used, shown in Figure \ref{phoneset1}. 
	Among the 20 teams that introduced their phone sets building methods, 11 teams use totally separate phone sets for Mandarin and English, 6 teams bind partial phones according to phonetics. 2 teams map all of the English phones to Chinese phones, above that, one team marked partial English words which appear frequently with Chinese phones.
	
	Concatenating a Chinese lexicon and an English lexicon is the most simple and commonly used method, and the merged phone sets cat be extracted from the lexicon directly. However, binding partial Chinese and an English phonemes is proved to obtain improvement by teams with higher ranking. The last method `map partial English words' refers to marking the high frequency English words with Chinese phonemes, with the rest part of the lexicon still using the first method. However, there is no detailed contrastive experiment results about phone sets reported.
	
%	\vspace{-0.3cm}
%	
%	\begin{figure}[htbp]
%		\centering
%		
%		\includegraphics[width=\linewidth,scale=1.00]{figure/pie}
%		\caption{The proportion of four methods for constructing mixed phoneme sets from descriptions of 20 teams.}
%		\label{phoneset1}
%	\end{figure}
%	\vspace{-0.6cm}
	\subsubsection{Feature Extraction and Data Augmentation}
	
%	\noindent\textbf{Feature Extraction and Data Augmentation: }
	Concatenating i-vectors feature to MFCC or Fbank feature brings 5\%-7\% relative improvement.
	Librispeech data is abandoned by most teams as it raised error rate, even when using only a small part of it. This may because of the mismatch of native English speaker and Chinese speaker.
	Speed augmentation can enhance the robustness of the model modestly, while volume augmentation and reverberation simulation help little. This may because the training data and test data are collected in the  environment with similar acoustic conditions.
	Spec-augment is a data augmentation method proposed by Google\cite{park2019specaugment}. Several teams gain about 2\% relative improvement using spec-augment layer in Kaldi Nnet3. 
	
	\subsubsection{Network Structure}
%	\noindent\textbf{Network Structure: }
	Kaldi chain model with lattice-free maximum mutual information (LF-MMI)\cite{Povey2016Purely} is used by all the teams, there are seldom differences among different systems. CNN or LSTM are used to combine with time-delay neural network (TDNN). The 1st team MobvoiASR used max-likelihood path to fix the original loss function and gain 3\% relative improvement. The 2nd place team Qdreamer use LF-MMI-SMBR (State-level Minimum Bayes Risk) and gain 9\% relative MER reduction comparing to original LF-MMI. 
	
	\begin{table}[hb] 
		
		\caption{The top 10 teams in track 2, along with their key method used, MER(\%) stands for mixture error rate. MERR are calculated based on their results in track 1. A negative MERR indicates that participants achieved a better LM than the 3-gram released in track 1.}
		\vspace{-0.08cm}
		\centering 
		\includegraphics[width=\linewidth,scale=1.00]{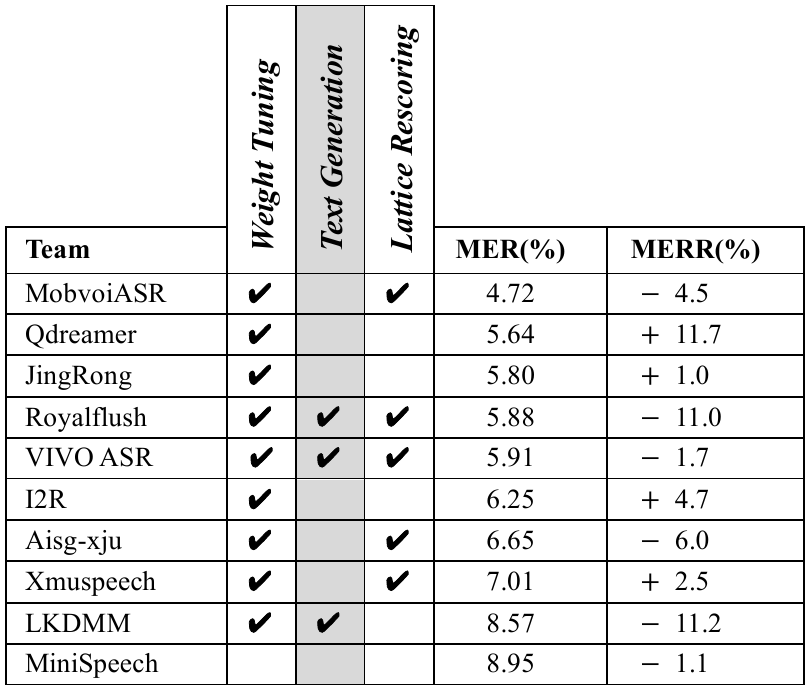} 
		\label{track2}
		\vspace{-0.9cm}
	\end{table}

\begin{table*}[!htbp] 
	\caption{Top 10 of submitted systems in track 3.}
	\centering 
	\vspace{-0.1cm}
	\includegraphics[width=\linewidth,scale=1.00]{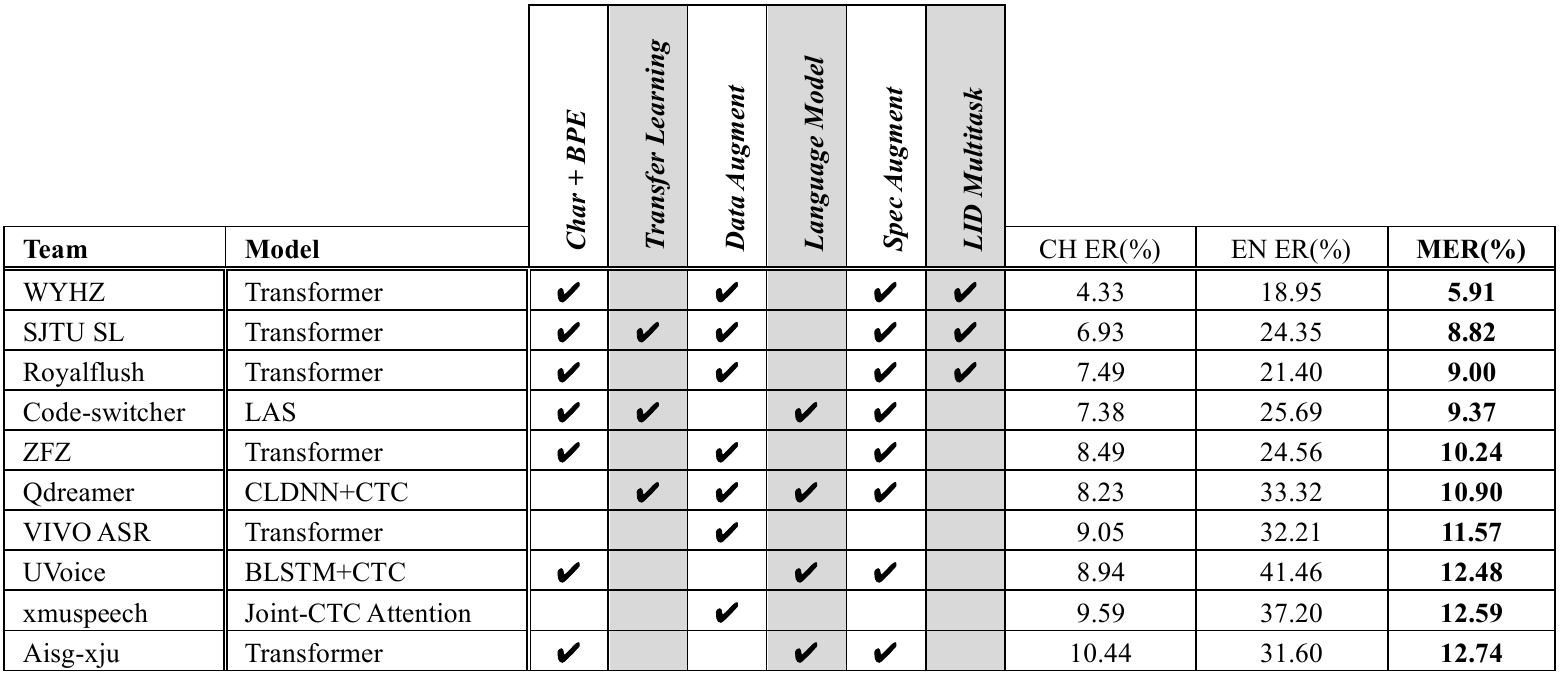} 
	\label{track3}
	\vspace{-0.75cm}
\end{table*}
	
	\subsection{Track2}
	
	The recognition results of the top 10 systems are shown in Table \ref{track2}. The efforts teams made for track 2 are mainly about CS text generation, balancing Chinese and English text proportion, and RNN-LM rescoring.
	
	Spontaneous code-switching text data for LM training is always scarce because of the randomness and casualness of CS phenomenon. Therefore, it's necessary to expand the text data. The parallel language pair in machine translation is widely used in ASR to generate CS transcripts. Besides, text generation based on pointer generator\cite{see2017get} is used by team Royalflush, but this method is limited by the scale of CS text and only a small amount of available data is generated.
	As reported, a well-trained RNN-LM using external CS data can yield 2\%-4\% recognition improvement.

%\vspace{-1.2cm}

	\subsection{Track3}
	
	In E2E track, 29 teams submit their results, the top 10 teams' results and key features are described in Table \ref{track3}. The outcomes of this track are mainly about modeling units, network structure, as well as taking language identification into consideration.

	\subsubsection{Modeling Units}
	
	The E2E ASR systems use sequence-to-sequence model to map the speech frames to the character sequence. In Chinese ASR, character is commonly used to be the modeling units directly as the amount of Chinese characters is around 6k. But modeling words in English directly is difficult because of the large amount and the sparsity of low frequency words. So in the challenge, Chinese character and English word piece\cite{chiu2018state} is mostly used. Its advantages mainly come from two aspects: balancing the granularity of Chinese and English modeling units and solving OOV problem with limited English training data. The number of English word pieces that teams used varied from 1k to 3k. Except for char + bpe, there are also teams using syllable for Chinese and letter for English.
	
	\subsubsection{Network and Language Modeling}
	
	The winner of track 3 went to a Transformer model\cite{seki2018end} trained by ESPnet\cite{watanabe2018espnet}, using multi-task learning to guide the decoder to distinguish Chinese and English characters (as reported, the language distinguishing CE loss optimized at decoder outperformed it at encoder). 
	Label smoothing, averaging checkpoints and spec-augment all yield recognition improvements. Data augmentation performs almost same as in track 1. 
	Transfer learning in the table refers to all kinds of different languages pre-training and fine-tuning strategies.
	
	\subsubsection{Language Modeling}
	
	As for language information modeling, 4 of the 10 teams in Table \ref{track3} use language model for rescoring or fusion with AM. Aisg-xju's and Royalflush's language models are RNN-LMs used for shallow fusion. UVoive's language model is a 4-gram LM used in CTC prefix beam search.  Qdreamer uses a 3-gram LM as first pass and an RNN-LM for rescoring.

	\section{Conclusions}
	
	In the ASRU 2019 code-switching automatic speech recognition challenge, participants used 500 hours Mandarin speech data and 200 hours intra-sentencial CS data to build ASR systems with recognition ability for Mandarin and English within a single utterance. Most teams achieve 5\% Chinese part error rate and English error rate under 20\% with DNN-HMM based models. The E2E models haven't outperformed the traditional model yet. It is clear that the systems tend to have higher recognition accuracy for Chinese part in the utterance, the reason may come from the imbalance of the data in two languages, which brings difficulty for LM training. The grammar of skipping between English words is completely invalid. 
	According to the results of the three tracks aforementioned, traditional ASR trained by Kaldi chain model outperformed the E2E models, but the gap is quickly narrowing. The result has highlighted that the detail of pronunciation lexicon and neural network effect a lot. In track 2, text generation is proved to be the most effective way to augment the language model, both word substitution according to grammatical rules and generative neural network help in data expansion. It is worth noting that RNN-LM did not replace N-gram LM but complemented it. As to E2E models, it turns out that attention based models performed more competitive and language identification help the model distinguish languages. Besides, spec-augment is proved a robust method of data augmentation with obvious performance gain.
	
%	Though attention-based models like Transformer have gained such success, deploying it for streaming speech recognition still remains a difficult task. Many researches aim at this aspect. Triggered attention\cite{moritz2019triggered}, chunk-wise attention encoder/decoder\cite{miao2020transformer}, time-restricted self-attention\cite{moritz2020streaming} are applied in attention-based models and get noteworthy results. 
	%In code-switch feild, several researches were about regularizing the distribution of parameters for different languages
	
	In this challenge, however, only recognition accuracy is considered in the evaluation. In the future, higher and more comprehensive requirements will be put forward, like streaming ASR system and ASR under complex acoustic environments.

	%\section{Acknowledgements}		

\bibliographystyle{IEEEtran}

\bibliography{mybib}

\end{document}